# NONLINEAR ELECTRON HEAT CONDUCTION EQUATION AND SELF SIMILAR METHOD FOR 1-D THERMAL WAVES IN LASER HEATING OF SOLID DENSITY DT FUEL

A. Mohammadian Pourtalari[1], M. A. Jafarizadeh[2], and M. Ghoranneviss[1]

[1]Plasma Physics Research Centre, Science and Research Branch,
Islamic Azad University, PO Box 14665-678, Tehran, Iran
[2]Department of Theoretical Physics and Astrophysics, Tabriz University,
Tabriz 51664, Iran

Electron heat conduction is one of the ways that energy transports in laser heating of fusible target material. The aim of Inertial Confinement Fusion (ICF) is to show that the thermal conductivity is strongly dependent on temperature and the equation of electron heat conduction is a nonlinear equation. In this article, we solve the one-dimensional (1-D) nonlinear electron heat conduction equation with a self-similar method (SSM). This solution has been used to investigate the propagation of 1-D thermal wave from a deuterium-tritium (DT) plane source which occurs when a giant laser pulse impinges onto a DT solid target. It corresponds to the physical problem of rapid heating of a boundary layer of material in which the energy of laser pulse is released in a finite initial thickness.

Address correspondence to Alireza Mohammadian Pourtalari, Plasma Physics Research Centre, Science and Research Branch, Islamic Azad University, Tehran, Iran. E-mail: amp_pprc@yahoo.com

# 1. INTRODUCTION

If the laser energy is released within the plasma and heated to a sufficiently high temperature, then a heat flux transported by heat conduction will appear. In most laser-produced plasmas, the hydrodynamic energy transport substantially exceeds the heat conduction. However, in the interaction of very short laser pulses (of the order of picoseconds or less) with material, the hydrodynamic energy transporting does not have any time for developing and therefore in these cases the heat conduction is dominant [1]. For temperatures which are not too high, it is in the form of ordinary (linear) heat conduction which serves as the mechanism of heat transfer. Radiation (nonlinear) heat conduction is the other heat transfer mechanism that comes into play at temperatures of the order of tens and hundreds of thousands of degree. The essential difference between linear and nonlinear heat conduction processes lies in the fact that the coefficient of nonlinear heat conductivity in the plasma is strongly temperature dependent [2]. Nonlinear heat conduction in the early phases of rapid laser energy input to material is a classical problem in the field of high-temperature hydrodynamics and it has an important role in thermonuclear reactions. Electrons or X-rays may serve as heat carriers [3, 4]. With a fusible target material, thermonuclear reactions will occur in the plasma, and reactions from such plasmas have been detected [5]. Because of unequal heating in reaction products, electrons and ions in the plasma are at different temperatures. Owing to the smallness of the electron mass in compared with the ion mass, electrons have a much higher conductivity than the ions. The thermal conductivity is nonlinear being a strong function of the temperature of the electrons. Modern high-power lasers can generate extreme states of matter that are relevant to fusion energy research [6, 7].

The interaction of an intense laser beam with a plane solid target has been considered by a number of authors [8-10]. In the last decades, the dominant role of heat conduction in the early stages of laser interaction with solid targets has been widely appreciated [11-17]. The mechanisms for heat transport in high temperature laser produced plasmas have been a topic of extensive research. The importance of nonlocal heat transport effects due to hot electrons in these plasmas was identified some years ago and a substantial amount of theoretic work has been conducted to understand its consequence [18-21]. Inertial confinement fusion (ICF) is an approach to fusion that relies on the inertia of the fuel mass to provide confinement. In the classical scheme of ICF, a spherical capsule containing a few milligrams of DT is directly irradiated by laser beams [22, 23]. A capsule containing thermonuclear fuel is compressed in an implosion process to conditions of high density and temperature. In direct-drive scheme the laser light propagates up to the critical density and heats an external shell of the capsule which expands. In recent years, application of the heat conduction problem in the physics of inertial confinement fusion has prompted significant renewed interest. Since laser pulses of high power in the range of picoseconds are available, the scheme of fast ignition [24] is studied as a possible approach for generating fusion energy from the thermonuclear reaction of DT.

In general, the inclusion of nonlinearity in heat conduction equation greatly restricts the number of analytical solutions which can be found. It is possible to obtain exact solutions for the heat conduction in some special cases. Similarity solutions were examined in [25-28]. The mathematics of similarity solutions has been extensively developed in the literature.

The main object of this research is presenting a solution based on SSM for nonlinear electron heat conduction equation in which the released energy of laser pulse at the fusible target material boundary makes a finite temperature for the electrons at the initial instant of time $(t=0)$. Results can be useful for understanding the electron temperature in a 1-D thermal wave, velocity of propagation of 1-D thermal wave, position of the heat front, heat flux, and heating domain in laser heating of solid density DT fuel.

This paper is organized as follows: In Section 2, some of the published works in other textbooks and journal articles about the solution of nonlinear heat conduction equation has been reviewed. Section 3 is devoted to the fundamental equations of the problem based on a hydrodynamic model. Mathematical methodology for the presentation of a self-similar solution for nonlinear electron heat conduction equation and results are presented in section 4. At the end of the paper, a brief conclusion will be presented.

# NOMENCLATURE

| | | | |
|---|---|---|---|
| $T$ | temperature | $\mu_e$ | electron viscosity coefficient |
| $\chi$ | coefficient of thermal diffusivity | $\tau_{ei}$ | equilibration time |
| $a$ | constant | $E_\alpha$ | energy of the $\alpha$ particles |
| $Q$ | energy | $f$ | fraction of absorbed energy |
| $f(\xi)$ | dimensional function | $L_0$ | initial length scale |
| $\xi$ | similarity variable | $T_{0e}$ | electron initial temperature |
| $\xi_0$ | constant | $g(t)$ | function of time |
| $\rho$ | mass density | $h(\xi)$ | function of similarity variable |
| $u$ | mass velocity | $\Theta$ | function of time |
| $Y$ | fraction of material burned | $\psi$ | constant in SSM |
| $W$ | reaction rate function | $\eta$ | constant in SSM |
| $n_\alpha$ | $\alpha$ particle density | $\varphi$ | trial function |
| $n_n$ | neutron density | $\beta$ | dimensionless similarity parameter |
| $n_D$ | deuterium density | $\zeta$ | eigenvalue |
| $n_T$ | tritium density | $x_f$ | front coordinate of thermal wave |
| $\sigma$ | fusion cross section | $E_{in}$ | input energy |
| $v$ | relative velocity of the two nuclei | $u_{hw}$ | thermal wave velocity |
| $T_e$ | electron temperature | $\tau_c$ | conduction time |
| $T_i$ | ion temperature | $\tau$ | laser pulse duration |
| $K_e$ | electron thermal conductivity coefficient | $q_H$ | heat flux |

## 2. PRELIMINARIES

In this section, we briefly present some of the solutions and properties related to the heat conduction equation. The heat conduction equation in the 1-D geometry can be written as:

$$\frac{\partial T}{\partial t} = \frac{\partial}{\partial x} \chi \frac{\partial T}{\partial x} \qquad (1)$$

The coefficient of thermal diffusivity is defined by:

$$\chi = aT^n \qquad (2)$$

If $n = 0$, i.e. a constant thermal diffusivity, then equation (1) is the simple diffusion equation:

$$\frac{\partial T}{\partial t} = a \frac{\partial^2 T}{\partial x^2} \qquad (3)$$

Equation (3) is the linear heat conduction equation. The theory of "thermal waves" is generated out of the nonlinear dependence of conductivity on temperature. In plane symmetric case, the conduction of heat in a uniform medium, whose properties vary nonlinearly with temperature, is most simply studied by writing the equation of heat conduction in the form:

$$\frac{\partial T}{\partial t} = a \frac{\partial}{\partial x} (T^n \frac{\partial T}{\partial x}) \qquad (4)$$

The law of propagation of heat from a source can be obtained by estimating the order of magnitude of the characteristic dimension of the heated region, or from dimensional considerations. In the earlier published works, Zeldovich [2] described some of the analytic solutions for a thermal wave driven by an instantaneous deposition of energy at the material boundary.

Here, Zeldovich's solution is presented as an example of these solutions. The solution of nonlinear heat conduction equation has been given by Zeldovich as:

$$T = (\frac{Q^2}{at})^{1/(n+2)} f(\xi) \qquad (5)$$

where

$$f(\xi) = \left[\frac{n}{2(n+2)} \left(\xi_0^2 - \xi^2\right)\right]^{1/n} \qquad (6)$$

$\xi_0$ and $\xi$ are defined as:

$$\xi_0 = \left[\frac{(n+2)^{1+n} 2^{1-n}}{n\pi^{n/2}}\right]^{1/(n+2)} \left[\frac{\Gamma(1/2+1/n)}{\Gamma(1/n)}\right]^{n/(n+2)} \quad (7)$$

$$\xi = \frac{x}{(aQ^n t)^{1/(n+2)}} \quad (8)$$

This solution appears to be erroneous and gives an infinite temperature in the initial instant of time $t=0$ (i.e. the temperature is singular $(\to \infty)$). It is to be underlined that Zeldovich has solved the nonlinear heat conduction equation in a model of a mathematical problem. Mathematically, the solution can be infinite, but using this solution in the physical problem of heat transfer by hot electrons, without correction of initial boundary conditions and using appropriate boundary conditions which agrees with physical concepts, is considered as a mistake. Physically, a solution should be realistic. Therefore, using this solution in the physical problems of propagation of thermal waves in the thermonuclear reactions is a wrong.

Chu's solution [11] is mentioned as an example of using Zeldovich's solution in thermonuclear reactions. Chu has given the solution of nonlinear heat conduction equation, as fallow:

$$T_e = \left(\frac{Q^2}{at}\right)^{2/9} f(\xi) \quad (9)$$

with

$$f(\xi) = \left[\frac{5}{18}(\xi_0^2 - \xi^2)\right]^{2/5} \quad (10)$$

where

$$\xi_0^{9/2} = \frac{(9/2)^{7/2} \, 2^{-3/2}}{\frac{5}{2}\pi^{5/4}} \left[\frac{\Gamma(9/10)}{\Gamma(2/5)}\right]^{5/2} \quad (11)$$

$$\xi = \frac{x}{(Q^{5/2} at)^{2/9}} \quad (12)$$

We revised the mathematical solution of Zeldovich based on a new SSM, and extend it at laser heating of solid density DT fuel.

## 3. REVISED HYDRODYNAMIC MODEL OF CHU

In order to see the importance of the correction of initial boundary conditions in the hydrodynamic equations, first the results of Chu [11] are going to be reproduced with a minimum of changes in used initial boundary conditions. It is to be underlined that the revised boundary condition agrees with the physical concepts, because it creates a finite electron temperature at the boundary layer the initial instant of time $(t=0)$.

For yielding the equation of mass conservation, the equations of continuity and thermonuclear reaction $(D+T \to \alpha + n)$ may be combined as fallow:

$$\frac{\partial \rho}{\partial t} + \frac{\partial}{\partial x}(\rho u) = 0 \tag{13}$$

and

$$\frac{\partial Y}{\partial t} + u \frac{\partial Y}{\partial x} = W \tag{14}$$

where $Y$ and $W$ defined by:

$$Y = \frac{n_\alpha + n_n}{n_D + n_T + n_\alpha + n_n} \tag{15}$$

and

$$W = \frac{1}{2} n (1-Y)^2 <\sigma v> \tag{16}$$

In the equation for $Y$, the $n$'s are the particle densities, and the subscripts describe the different particle species. But, in the equation for $W$, the $n$ stands for the total number density of the ions. Assuming that the plasma consists of deuterons and tritons of density $\frac{n}{2}$ each $(DT \approx 50\% - 50\%)$, the rate of fusion processes $W$ is such a hot dense plasma state is given by:

$$W = \frac{n^2}{4} <\sigma v>_{DT} \tag{17}$$

with

$$<\sigma v>_{DT} = 3.7 \times 10^{-12} T_i^{-2/3} \exp(-20 T_i^{-1/3}) \tag{18}$$

The electron temperature equations are expressing the conservation of energy:

$$\frac{\partial T_e}{\partial t} + u \frac{\partial T_e}{\partial x} = -\frac{2}{3} T_e \frac{\partial u}{\partial x} + \frac{2 m_i}{3 k_b \rho} \mu_e \left(\frac{\partial u}{\partial x}\right)^2 + \frac{2 m_i}{3 k_b \rho} \frac{\partial}{\partial x} \left(K_e \frac{\partial T_e}{\partial x}\right) + W_e + \frac{T_i - T_e}{\tau_{ei}} - A \rho T_e^{\frac{1}{2}} \quad (19)$$

where

$$\mu_e = \frac{0.406 \, m_e^{1/2} (k_b T_e)^{5/2}}{e^4 \, Ln\Lambda} \quad (20)$$

$$K_e = 1.89 \times \left(\frac{2}{\pi}\right)^{3/2} \left(\frac{k_b^{7/2} T_e^{5/2}}{m_e^{1/2} e^4 \, Ln\Lambda}\right) \quad (21)$$

$$\tau_{ei} = \frac{3 m_i m_e \, k_b^{3/2}}{8 \times (2\pi)^{1/2} n e^4 \, Ln\Lambda} \left(\frac{T_i}{m_i} + \frac{T_e}{m_e}\right)^{3/2} \quad (22)$$

The energy transfer term $W_e$ is related to $W$ by:

$$W_e = (1 - f) E_\alpha W \quad (23)$$

where

$$E_\alpha = 3.5 \; [MeV] \approx 0.5607 \times 10^{-5} \; [erg] \quad (24)$$

The function $f$, which is the fraction of $\alpha$ particle energy absorbed by the ions, is given by:

$$f = \left(1 + \frac{32}{T_e}\right)^{-1} \quad [T_e \; in \; keV] \quad (25)$$

The last term on the right-hand side of equation (19) is the bremsstrahlung term. $A$ is a constant taken for pure bremsstrahlung with $Z = 1$:

$$A = \left(\frac{2}{3 k_b m_i}\right) \times 1.42 \times 10^{-27} \; [°K^{1/2}.cm^3.g^{-1}.\sec^{-1}] \quad (26)$$

The initial boundary conditions are defined by:

$$\rho(x,t)|_{t=0} = \rho_0$$

$$u(x,t)|_{t=0} = 0$$

$$Y(x,t)|_{t=0} = 0$$

$$T_e(x,t)|_{t=0} \neq \delta(x) \quad or \quad \infty \quad \rightarrow \quad (finite \quad electron \quad temperature)$$

$$W(\rho, Y, T_e)|_{t=0} = 0$$

$$u(x,t)|_{x=0} = 0$$

$$\frac{\partial T_e}{\partial x}(x,t)|_{x=0} = 0 \tag{27}$$

Hence, equation (19) reduces to:

$$\frac{\partial T_e}{\partial t} = \frac{2m_i}{3k_b \rho} \frac{\partial}{\partial x}(K_e \frac{\partial T_e}{\partial x}) - \frac{T_e}{\tau_{ei}} - A\rho T_e^{1/2} \tag{28}$$

In equation (28) the conduction term involves two spatial differentiations. The other two terms, equilibration and bremsstrahlung are small in compared with the conduction term. By neglecting these smaller terms, the resulting equation can be rewritten as:

$$\frac{\partial T_e}{\partial t} = a \frac{\partial}{\partial x}(T_e^{5/2} \frac{\partial T_e}{\partial x}) \tag{29}$$

where

$$a = (\frac{2m_i}{3k_b \rho}) \times 1.89 \times (\frac{2}{\pi})^{3/2} \times \frac{k_b^{7/2}}{m_e^{1/2} e^4 \, Ln \Lambda} \tag{30}$$

In Table 1, the values of ion mass, electron mass, Boltzmann constant, initial density, electron charge, and Spitzer logarithm are presented:

**Table1.** Values of constant parameters

| parameters | value |
| --- | --- |
| ion mass | $m_i = 1.6726 \times 10^{-24} [g]$ |
| electron mass | $m_e = 9.1091 \times 10^{-28} [g]$ |
| Boltzmann constant | $k_b = 1.3806 \times 10^{-16} [erg.°K^{-1}]$ |
| initial density | $\rho = 0.1964 [g.cm^{-3}]$ |
| electron charge | $e = 4.8030 \times 10^{-10} [cgs] [e$ in |
| Spitzer logarithm | $Ln\Lambda \approx 8.9$ |

## 4. MATHEMATICAL METHODOLOGY AND RESULTS

We now present a solution for nonlinear electron heat conduction equation based on a new SSM. Similarity solution to equation (29) in 1-D geometry can be found by assuming a product solution with the fallowing form:

$$T_e(x,t) = T_{0e} g(t) h(\xi) \qquad (31)$$

where $g(0) = h(0) = 1$.

The similarity variable is defined as:

$$\xi = \frac{x}{\Theta(t) L_0} \qquad (32)$$

Here, $L_0$ is the initial length scale (equal to Neodymium-glass laser wavelength $\approx 1.06 \times 10^{-4} cm$).

Notice that $g(t)$ describes the temperature dependence at the boundary (electron initial temperature $T_{0e}$).

Using equations (31) and (32) in equation (29), we obtain:

$$T_{0e}(\dot{g}h - gh'\xi \frac{\dot{\Theta}}{\Theta}) = \frac{5aT_{0e}^{7/2} g^{7/2} h^{3/2} h'^2}{2\Theta^2 L_0^2} + \frac{aT_{0e}^{7/2} g^{7/2} h^{5/2} h''}{\Theta^2 L_0^2} \qquad (33)$$

where the dots are time derivatives and the prime is a derivative with respect to $\xi$. If $\dot{\Theta} = -\frac{\Theta}{\xi}\frac{d\xi}{dt} \neq 0$, we can rewrite equation (33) as:

$$(\frac{\Theta\dot{\Theta}}{g^{5/2}})(\frac{\dot{g}\Theta}{g\dot{\Theta}} - \frac{\xi h'}{h}) = \frac{aT_{0e}^{5/2}}{L_0^2}(h^{5/2}h')'h^{-1} \tag{34}$$

For equation (34) to be satisfied, we consider $\frac{\dot{g}\Theta}{g\dot{\Theta}} \equiv \psi$ and $\frac{\Theta\dot{\Theta}}{g^{5/2}} \equiv \eta$ for a pair of constants $\psi$ and $\eta$. It is important to notice that the parameters $\psi$ and $\eta$ are restricted by the nature of the physical problem.

Choose the solution $g = \Theta^\psi$ for equation of $\psi$. Thus, $\Theta(0) = 1$ since $g(0) = 1$. Substitute this result in equation of $\eta$ to obtain the fallowing term:

$$\Theta\dot{\Theta} = \Theta^{(\frac{5}{2})\psi}\eta \tag{35}$$

then

$$\Theta(t) = \begin{cases} \left[1 + (2 - (\frac{5}{2})\psi)\eta t\right]^{1/(2-(5/2)\psi)} & \text{for} \quad (\frac{5}{2})\psi \neq 2 \\ e^{\eta t} & \text{for} \quad (\frac{5}{2})\psi = 2 \end{cases} \tag{36}$$

where we have satisfied the boundary condition $\Theta(0) = 1$.

Thus, equation (36) indicates that for having a forward-going thermal wave, we must have $\eta > 0$. For avoiding any infinite velocities, the domain of $\psi$ is restricted to $\psi \leq \frac{4}{5}$. With the above transformations and initial scale length defined by $L_0^2 = aT_{0e}^{5/2}\eta^{-1}$, we are left with the differential equation in the similarity variable $\xi$:

$$h\psi - h'\xi = \frac{5}{2}h^{3/2}h'^2 + h^{5/2}h'' \tag{37}$$

The boundary conditions are $h(0) = 1$ and the requirement that the heat flux must go to zero inside the material at some point of $\xi_0 > 0$.

Although standard numerical techniques can be readily applied to the solution of equation (37), it is instructive to obtain a self-similar solution. We choose a trial function $\varphi$ given by:

$$\varphi = 1 - \beta^2 \tag{38}$$

We let $\beta = \xi / \xi_0$ and $h = \varphi^{2/5}$, such that $h(\xi_0) = \varphi^{2/5}(1) = 0$. Thus:

$$h' = \frac{2}{5\xi_0} \varphi^{-3/5} \varphi' \tag{39}$$

and

$$h'' = \frac{2}{5\xi_0^2} \left[ (-\frac{3}{5}) \varphi^{-8/5} \varphi'^2 + \varphi^{-3/5} \varphi'' \right] \tag{40}$$

By substituting $h$, $h'$ and $h''$ into equation (37), we have:

$$\zeta \left[ (\frac{5}{2}) \psi \varphi - \beta \varphi' \right] = (\frac{5}{2}) \varphi \varphi'' + \varphi'^2 \tag{41}$$

where $\zeta = \frac{5}{2} \xi_0^2$, and the prime is a derivative with respect to $\beta$.

The equation (41) is now cast in the form of a nonlinear eigenvalue problem, where we must find the positive value of $\zeta$ such that $\varphi$ gets monotone and decrease on the interval $0 \le \beta \le 1$ and then goes to zero at $\beta = 1$. This equation has an exact solution in the special case, where it is corresponds with the Zeldovich's solution [2]. This case corresponds to the physical problem of rapid (compared with the thermal wave motion) heating of a boundary layer of material which drives a thermal wave with the average temperature dropping as more material is heated.

In our solution, the energy is released in a finite initial thickness of dimension $x_f(0) = \xi_0 L_0$. In this case, we have to choose $\psi = -1$ and $\zeta = 2$, therefore:

$$\Theta(t) = \left[ 1 + (\frac{9}{2}) \eta t \right]^{2/9} \tag{42}$$

and

$$g(t) = \frac{1}{\Theta(t)} \tag{43}$$

By substituting $h = \varphi^{2/5}$ in equation (38), we have:

$$h(\xi) = (1-\beta^2)^{2/5} = (1 - \frac{\xi^2}{\xi_0^2})^{2/5} \tag{44}$$

By substituting $\xi$ from equation (32) and $\xi_0$ in the form $\xi_0 = \dfrac{x_f(t)}{\Theta(t)L_0}$ into equation (44), we have:

$$h(\xi) = \left[1 - \frac{x^2}{x_f^2(t)}\right]^{2/5} \tag{45}$$

## 4.1. Electron temperature

By substituting $g(t)$ from equation (43) and $h(\xi)$ from equation (45) into equation (31), the final solution of equation (29) is given by:

$$T_e(x,t) = \frac{T_{0e}}{\Theta(t)}\left[1 - \frac{x^2}{x_f^2(t)}\right]^{2/5} \tag{46}$$

Here, the energy is released in a finite initial thickness of dimension:

$$x_f(t) = (\frac{4}{5})^{1/2} L_0 \left[1 + (\frac{9}{2})\frac{aT_0^{5/2}}{L_0^2}t\right]^{2/9} \tag{47}$$

By substituting $x_f(t)$ from equation (47) and $\Theta(t)$ from equation (42) into equation (46), we will have the fallowing term:

$$T_e(x,t) = \frac{T_{0e}}{\left[1 + (\frac{9}{2})\frac{aT_{0e}^{5/2}}{L_0^2}t\right]^{2/9}} \left[1 - \frac{x^2}{\frac{4}{5}L_0^2\left[1 + (\frac{9}{2})\frac{aT_{0e}^{5/2}}{L_0^2}t\right]^{4/9}}\right]^{2/5} \tag{48}$$

The electron initial temperature $T_{0e}$ is determined by using the energy conservation at the initial instant of time $(t=0)$:

$$Q = \int_{x=-(\frac{4}{5})^{1/2}L_0}^{x=(\frac{4}{5})^{1/2}L_0} T_e(x,0)dx \qquad (49)$$

At the initial instant of time $(t=0)$, equations (47) and (48) give:

$$x_f(0) = (\frac{4}{5})^{1/2} L_0 \qquad (50)$$

and

$$T_e(x,0) = T_{0e}\left[1 - \frac{5x^2}{4L_0^2}\right]^{2/5} \qquad (51)$$

Therefore, at the other instants of time $(t>0)$, the heat wave propagates into both directions away from the plane $x_f(0)$. By substituting $T_e(x,0)$ in equation (51) for equation (49), and using of the fallowing integral form:

$$\int_0^a (a^m - x^m)^p dx = \frac{a^{1+mp}\Gamma(1/m)\Gamma(p+1)}{m\Gamma[(1/m)+p+1]} \qquad (52)$$

For $m=2$, $p=\frac{2}{5}$ the solution $Q$ is:

$$Q = T_{0e}(\frac{2L_0}{\sqrt{5}})\left[\frac{\Gamma(1/2)\Gamma(7/5)}{\Gamma(19/10)}\right] \qquad (53)$$

then

$$T_{0e} = \frac{\sqrt{5}}{2L_0}Q\left[\frac{\Gamma(19/10)}{\Gamma(1/2)\Gamma(7/5)}\right] \qquad (54)$$

with

$$Q = \frac{E_{in}}{(\frac{3\rho k_b}{2m_i})} \qquad (55)$$

At the times $t \gg (\frac{2}{9})\frac{L_0^2}{aT_{0e}^{5/2}}$, equation (48) gives:

$$T_e = (\frac{9}{2})^{-2/9} a^{-2/9} T_{0e}^{4/9} L_0^{4/9} t^{-2/9} \left[ 1 - \frac{x^2}{(\frac{4}{5})(\frac{9}{2})^{4/9} a^{4/9} T_{0e}^{10/9} L_0^{10/9} t^{4/9}} \right]^{2/5} \quad (56)$$

for the arbitrary energy input $E_{in} = 7.5 \times 10^{15} \, erg/cm^2$, we have:

$$T_e \approx 2.6787 \times 10^6 t^{-2/9} \quad (57)$$

Figure 1 shows typical computed electron temperature profiles at different instants of time $(t > 0)$:

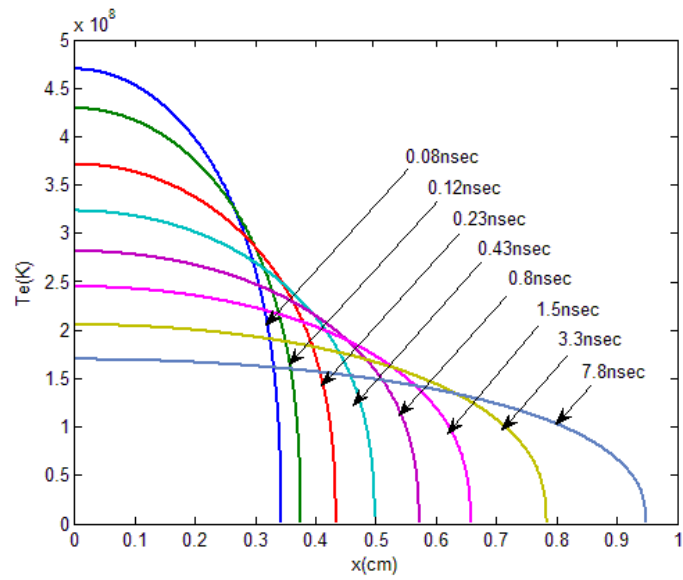

**Figure 1.** Computed electron temperature profiles at different instants of time

We believe that the solution of Chu based on Zeldovich's similarity solution is wrong because this solution creates an infinite temperature for the electrons $(T_e \to \infty)$ at the initial instant of time $(t=0)$. This can be seen in Figure 2, which depicts Chu's results of the computed electron and ion temperatures at different instants of time $(t>0)$ in irradiated solid density DT plane target:

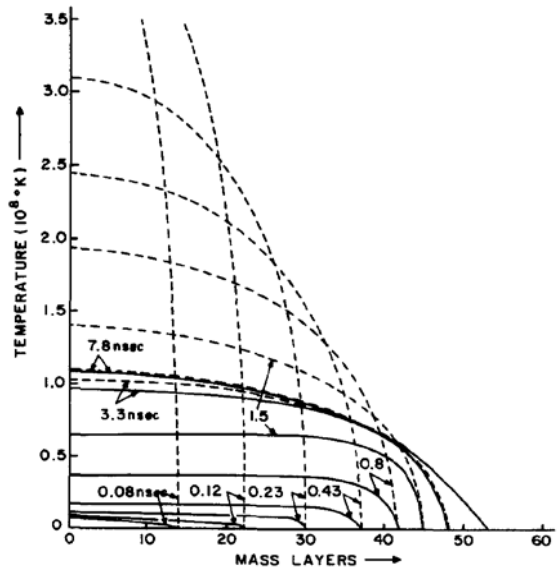

**Figure 2.** Characteristics of the electron temperature profiles at different instants of time from Figure 2 of Chu [11]

In this Figure, electron temperature curve increases infinitely for $(t<0.08 \; n\sec)$ and it goes to infinity at $(t=0)$ that physically it is meaningless. But, our solution gives a finite temperature for the electrons. Figure 3 shows typical computed electron temperature profiles for the arbitrary energy input $E_{in} = 7.5 \times 10^{15} \, erg/cm^2$ at the initial instant of time $(t=0)$:

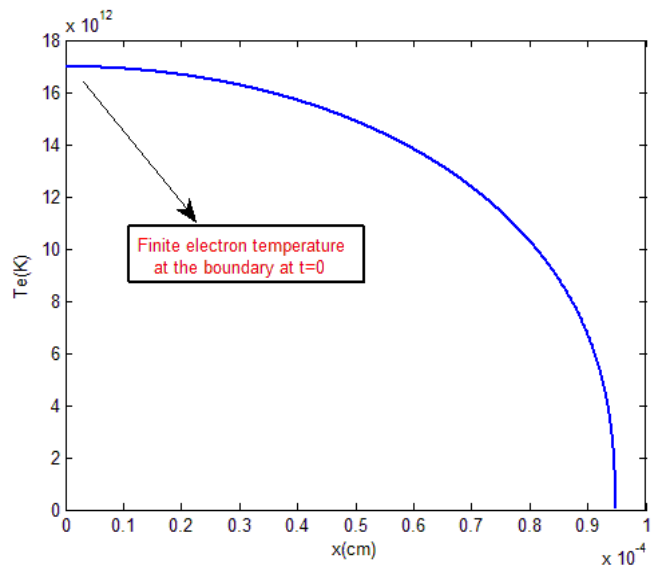

**Figure 3.** Computed electron temperature profile at the initial instant of time

## 4.2. Velocity of propagation of 1-D thermal wave

At the times $t \gg (\frac{2}{9})\frac{L_0^2}{aT_{0e}^{5/2}}$, equation (47) gives:

$$x_f(t) = (\frac{4}{5})^{1/2}(\frac{9}{2})^{2/9} a^{2/9} T_{0e}^{5/9} L_0^{5/9} t^{2/9} \tag{58}$$

The velocity of propagation of 1-D thermal wave $u_{hw}$ is defined by:

$$u_{hw} = \frac{dx_f}{dt} \tag{59}$$

From equation (58), the velocity of the heat front is given by:

$$u_{hw} = \dot{x}_f(t) = (\frac{4}{5})^{1/2}(\frac{9}{2})^{-7/9} a^{2/9} T_{0e}^{5/9} L_0^{5/9} t^{-7/9} \tag{60}$$

From equations (56) and (60), it appears that at the initial instant of time $(t=0)$ both the temperature and the thermal wave velocity are singular $(\to \infty)$. However, as was previously explained, equation (58) is relevant only for times $t \gg \tau$, where $\tau = (\frac{2}{9})\frac{L_0^2}{aT_{0e}^{5/2}}$. Therefore, it might be convenient to write the solutions for $T_e$ and $u_{hw}$, for times $t \gg \tau$, in the form:

$$\frac{T_e(t)}{T_e(\tau)} = (\frac{\tau}{t})^{2/9} \tag{61}$$

and

$$\frac{u_{hw}(t)}{u_{hw}(\tau)} = (\frac{\tau}{t})^{7/9} \tag{62}$$

## 4.3. Heat flux

The heat flux is given by:

$$q_H(x,t) = -K_e \frac{\partial T_e}{\partial x} \tag{63}$$

By using equation (56) it is found that:

$$q_H = (\frac{15}{4})(\frac{9}{2})^{-4/9} k_b \rho \, m_i^{-1} T_{0e}^{43/18} L_0^{-10/9} a^{5/9} t^{-4/9} \left[1 - \frac{x^2}{(\frac{4}{5})(\frac{9}{2})^{4/9} a^{4/9} T_{0e}^{10/9} L_0^{10/9} t^{4/9}}\right]^{2/5} x \tag{64}$$

This solution states that at $t$, the effect of conduction for the arbitrary energy input $E_{in} = 7.5 \times 10^{15} \, erg/cm^2$, will have penetrated at distance:

$$x_f \approx 60.0851 \, t^{2/9} \tag{65}$$

On the other hand, for a definite amount of energy, and a definite length $x_f$, conduction effects will not propagate to $x_f$ until a time:

$$\tau_c \approx 9.8980 \times 10^{-9} \, x_f^{9/2} \tag{66}$$

where $\tau_c$ is the conduction time.

The heat flux $q_H$ increases linearly to a good approximation from the origin $x=0$ to almost the front of the heat wave and drops very fast to zero at $x = x_f$, as shown in Figure 4.

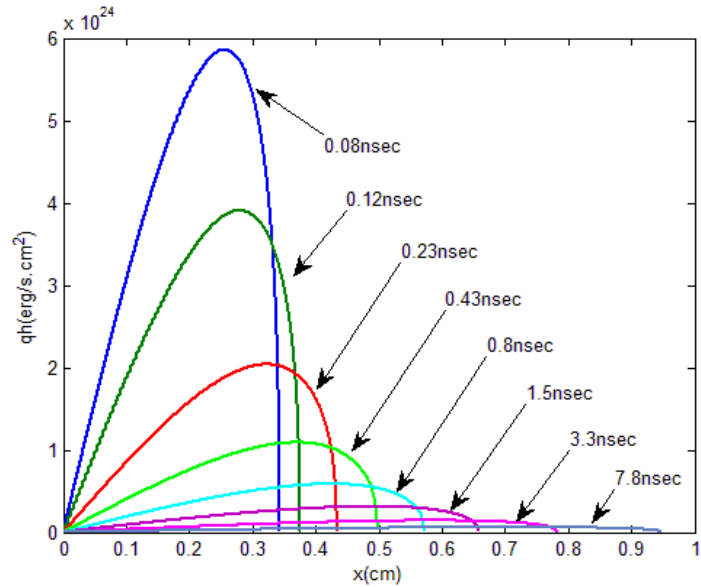

**Figure 4.** Distribution of heat flux in a 1-D thermal wave

## 4.4. Heating and cooling domains

The divergence of the heat flux is given by:

$$-\frac{\partial q_H}{\partial x} = K_e \frac{\partial}{\partial x}\left(\frac{\partial T_e}{\partial x}\right) \tag{67}$$

By using equation (56), it is found that

$$-\frac{\partial q_H}{\partial x} = -(\frac{15}{4})(\frac{9}{2})^{-4/9} k_b \rho \ m_i^{-1} a^{5/9} T_{0e}^{43/18} L_0^{-10/9} t^{-4/9}$$

$$\times \left[1 - \frac{x^2}{(\frac{4}{5})(\frac{9}{2})^{4/9} a^{4/9} T_{0e}^{10/9} L_0^{10/9} t^{4/9}}\right]^{-3/5} \left[1 - \frac{x^2}{(\frac{5}{9})(\frac{4}{5})(\frac{9}{2})^{4/9} a^{4/9} T_{0e}^{10/9} L_0^{10/9} t^{4/9}}\right] \quad (68)$$

The divergence of the heat flux is almost constant in the entire region of the plateau. The main region of hot plasma is cooled almost uniformly and only near the edge of the thermonuclear reaction wave in the plasma heated by energy removed from the main mass of DT plasma (sees Figure 5). The heat wave propagates in such a manner that the volume of DT plasma is cooled almost uniformly and the energy lost by the plasma is absorbed near the wave front that is the manner by which the wave continually encompasses the new layer of cold DT plasma.

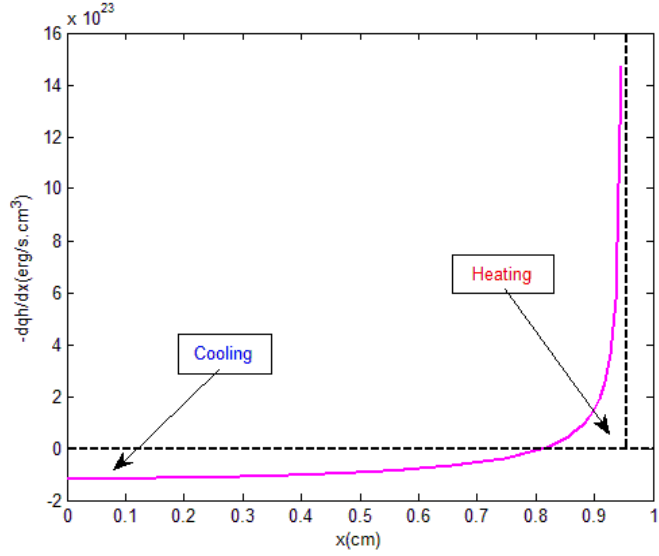

**Figure 5.** Divergence of heat flux in a 1-D thermal wave and cooling and heating domains

From the equation (68), one note that for $x < (\frac{4}{9})^{1/2} (\frac{9}{2})^{2/9} a^{2/9} T_{0e}^{5/9} L_0^{5/9} t^{2/9}$ the medium is cooled (i.e. the temperature decreases with time), while for $x > (\frac{4}{9})^{1/2} (\frac{9}{2})^{2/9} a^{2/9} T_{0e}^{5/9} L_0^{5/9} t^{2/9}$ the medium is heated. At $x = (\frac{4}{9})^{1/2} (\frac{9}{2})^{2/9} a^{2/9} T_{0e}^{5/9} L_0^{5/9} t^{2/9}$ there is a singularity in the increase (in time) in temperature. The main hot region is cooled almost uniformly and only the domain with the edge of the wave is heated very fast.

## 5. CONCLUSION

Since most of the problems related with hydrodynamic and plasma are more complex and analytically they rarity have an exact solution so the similarity methods are often useful in this field. If the required equations of the problem are certain, making dimensionless of equations can be used directly for obtaining the dimensionless parameters. In the current study problem, there are more related parameters which can affect on the model of physical problem. Therefore, we have tried to obtain the dimensionless parameters and study the problem in the view of dimension and dynamical similarity. Recognition of dimensionless parameters and self-similarity solution of nonlinear electron heat conduction equation has deepened our understanding from the phenomenon of electron heat conduction in DT plasma. In our solution, the released energy by laser pulse does not make infinite temperature for the electrons at the target material boundary at the initial instant of time $(t = 0)$. We expect that the possible useful results of this research can provide exact theoretical bases for doing further practical experiments in the field of laser-fusion.

# APPENDIX I. THE DEFLAGRATION MODEL

The basic flow structure associated with the one dimensional steady state deflagration model is shown in Figure 6. In this model, the laser radiation is considered to be absorbed in a narrow region, (the deflagration), which is in consequence heated generating a high pressure which drives the downstream expansion fan and an upstream shock wave.

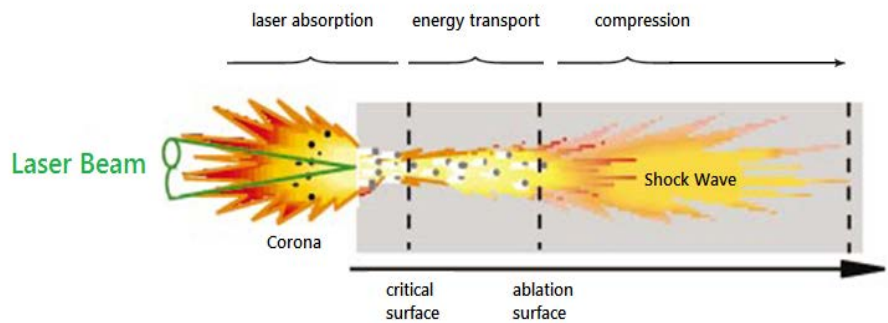

**Figure 6.** Basic flow structure associated with the one dimensional steady state deflagration model

# APPENDIX II. TYPICAL LASER IRRADIATION CONDITIONS IN DIRECT-DRIVE ICF

Inertial confinement fusion is a process where nuclear fusion reactions are initiated by heating and compressing a target that most often contains DT by the use of intense laser irradiations. The Schematic diagram illustrating the flow of energy from the laser to the target in direct-drive ICF is shown in Figure 7. Position of the layers is devoted to probe different regions of the target.

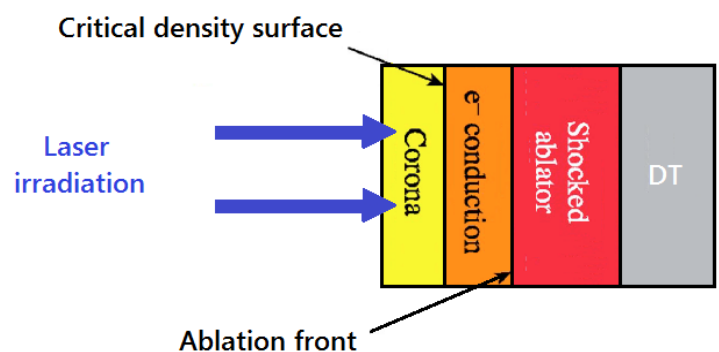

**Figure 7.** Schematic diagram illustrating the flow of energy from the laser to the target in direct-drive ICF

The laser energy is absorbed in the corona at densities less than the critical density via inverse bremsstrahlung. Thermal electron conduction transports the absorbed energy to the ablation front. Laser ablation launches a shock wave in the ablator or shell of the target.